\documentclass[prb]{revtex4}
\usepackage{graphicx}
\usepackage{bm}
\usepackage[latin1]{inputenc}
\def\ke{kinetic energy }
\def\etal{{\it et al.~}}
\begin{document}
\title{The most compact scission configuration of fragments from low energy fission of $^{234}$U and $^{236}$U }
\author{M. Montoya}
\email{mmontoya@ipen.gob.pe}
\affiliation{Instituto Peruano de Energ\'{\i}a Nuclear, Av. Canad\'a 1470, Lima 41, Per\'u.}
\affiliation{Facultad de Ciencias, Universidad Nacional de Ingenier\'{\i}a, Av. Tupac Amaru 210, Apartado 31-139, Lima, Per\'u.}
\date{September 18th, 2008 }
\begin{abstract}
Using a time of flight technique,  the maximal values of kinetic energy as a function of primary mass of fragments from low energy fission of $^{234}U$ and  $^{236}U$ were measured  by Signarbieux \etal From calculations of scission configurations, one can conclude that, for those two fissioning systems, the maximal value of total kinetic energy corresponding to fragmentations ($_{42}$Mo$_{62}$, $_{50}$Sn$_{80}$) and ($_{42}$Mo$_{64}$, $_{50}$Sn$_{80}$), respectively, are equal to the available energies, and their scission configurations are composed by a spherical heavy fragment and a prolate light fragment both in their ground state.

\textsl{Keywords}: Low energy fission; $^{234}U$; $^{236}U$; fragment kinetic energy; cold fission \\
PACS: 21.10.Gv; 25.85.Ec; 24.10.Lx
\\
\\

\end{abstract}

\maketitle
\section{Introduction}
\label{intro}

One of the most studied quantities to understand the fission process is the fragment mass and \ke distribution, which is very closely related to the topological features in the multi-dimensional potential energy surface~\cite{moller}. Structures on the distribution of mass and \ke may be interpreted by shell effects on potential energy of the fissioning system, determined by the Strutinsky prescription and discussed by Dickmann
\etal~\cite{dick} and Wilkins ~\etal\cite{wilkins}. 

In order to investigate the fragments with very low excitation energy, using the time of flight method, Signarbieux ~\etal\cite{Signarbieux} measured the fragment mas ($A$) distribution for high values of fragment kinetic energy. Because in that kinetic energy region there is no neutron emission, the time of flight technique permits separate neighboring fragment masses. In this work one calculates the deformations of those fragments which must correspond to the most compact scission configurations, i.e. to the highest values of Coulomb interaction energy between the two fragments.

\section{The most compact scission configurations}
\label{sec:compact}

In the process of thermal neutron induced fission of $^{233}U$, a composed nucleus $^{234}U^*$ with excitation energy equal to neutron separation energy ($\epsilon_n$) is formed first. Then, this nucleus splits in two complementary light and heavy fragments having $A_L$  and $A_H$ as mass numbers, and $E_L$ and $E_H$ as kinetic energies, respectively. 

The Q-value of the this reaction is given by the relation
\begin{equation}
Q (Z_L,A_L, Z_H, A_H) = M(92,234)- M(Z_L, A_L) - M(Z_H, A_H),
\label{eq:Q}
\end{equation}

where M(Z,A) is the mass of nucleus with $Z$ and $A$ as proton number and mass number, respectively.

This available energy at scission configuration is spend in prescission total kinetic energy ($TKE_0$), fragments interaction Coulomb energy, $CE$, fragments deformation energy,

\begin{equation}
TDE = DE_L + DE_H,
\label{eq:TDE}
\end{equation}

where ($DE_L$) and ($DE_H$) are the light and heavy fragment deformation energy, respectively, and in fragments intrinsic excitation energy,

\begin{equation}
TXE = XE_L + XE_H,
\label{eq:TXE}
\end{equation}

where ($XE_L$) and ($XE_H$) are the light and heavy fragment intrinsic excitation energy, respectively.

Then the balance energy at scission configuration results

\begin{equation}
Q + \epsilon_n = TKE_0 + CE + TDE + DXE.
\label{eq:QETDX}
\end{equation}

If there is no neutron emission, the light and heavy fragments reach the detectors with their primary kinetic energies equal to $KE_L$ and $KE_H$, respectively. The total primary fragments kinetic energy will be

\begin{equation}
TKE = KE_L + KE_H = TKE_0 + CE = Q + \epsilon_n - TDE - TXE.
\label{eq:TKEQDX}
\end{equation}

The maximal value of total kinetic energy is reached when the sum of TDE and TDX is minimal, i.e.

\begin{equation}
TKE_{max} =  (TKE_0 + CE)_{max} = Q + \epsilon_n - (TDE - TXE)_{min}.
\label{eq:TKEQDXmax}
\end{equation}

The most compact scission configuration occurs when maximal value of coulomb energy is equal to the available energy, i.e.

\begin{equation}
CE_{max} = Q + \epsilon_n.
\label{eq:CEmaxQ}
\end{equation}

In this case, from eq. ~\ref{eq:TKEQDX} one obtains the relations

\begin{equation}
TKE_{max} = CE_{max} = Q + \epsilon_n,
\label{eq:TKEmaxCQ}
\end{equation}

and

\begin{equation}
DE_{min} = 0, 
DX_{min} = 0, 
TKE_0 = 0.
\label{eq:DEDXmin}
\end{equation}

Not always this situation is possible to occur. Nevertheless we can assume that for each mass fragmentation the maximal value of total kinetic energy is obtained for similar condition, i.e. $TKE_0 = 0$, $TXE = 0$ and $TDE=TDE_{min}$.

\section{Deformation energy}

\begin{figure}
\centering
\includegraphics[height=0.5\textwidth]{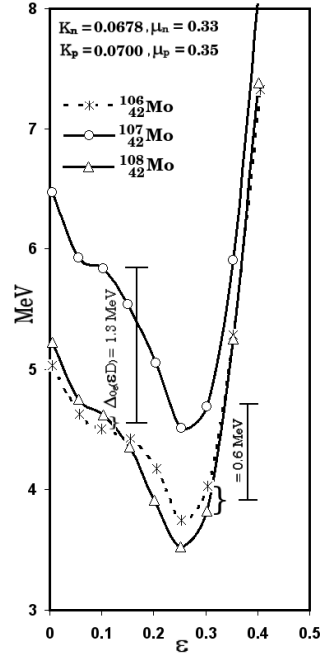}
\caption{Deformation energy for nuclei $^{106-108}$Mo calculated by a drop liquid model with pairing and shell correction~\cite{Myers}. See text.}
\label{fig:106108Mo}
\end{figure}

\begin{figure}
\centering
\includegraphics[height=0.5\textwidth]{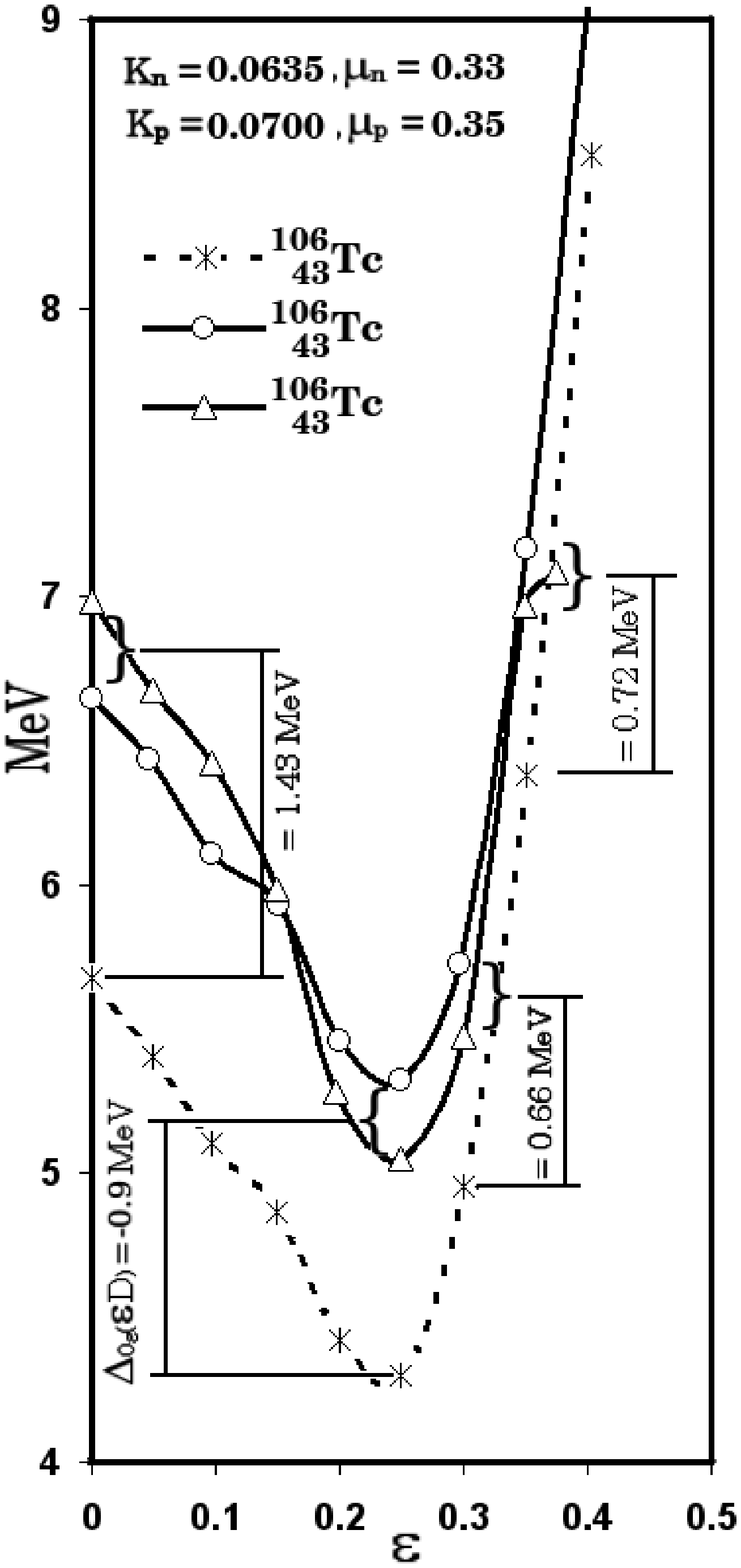}
\caption{Deformation energy for nuclei $^{106-108}$Tc calculated by a drop liquid model with pairing and shell correction~\cite{Myers}. See text.}
\label{fig:106108Tc}
\end{figure}

\begin{figure}
\centering
\includegraphics[height=0.45\textwidth]{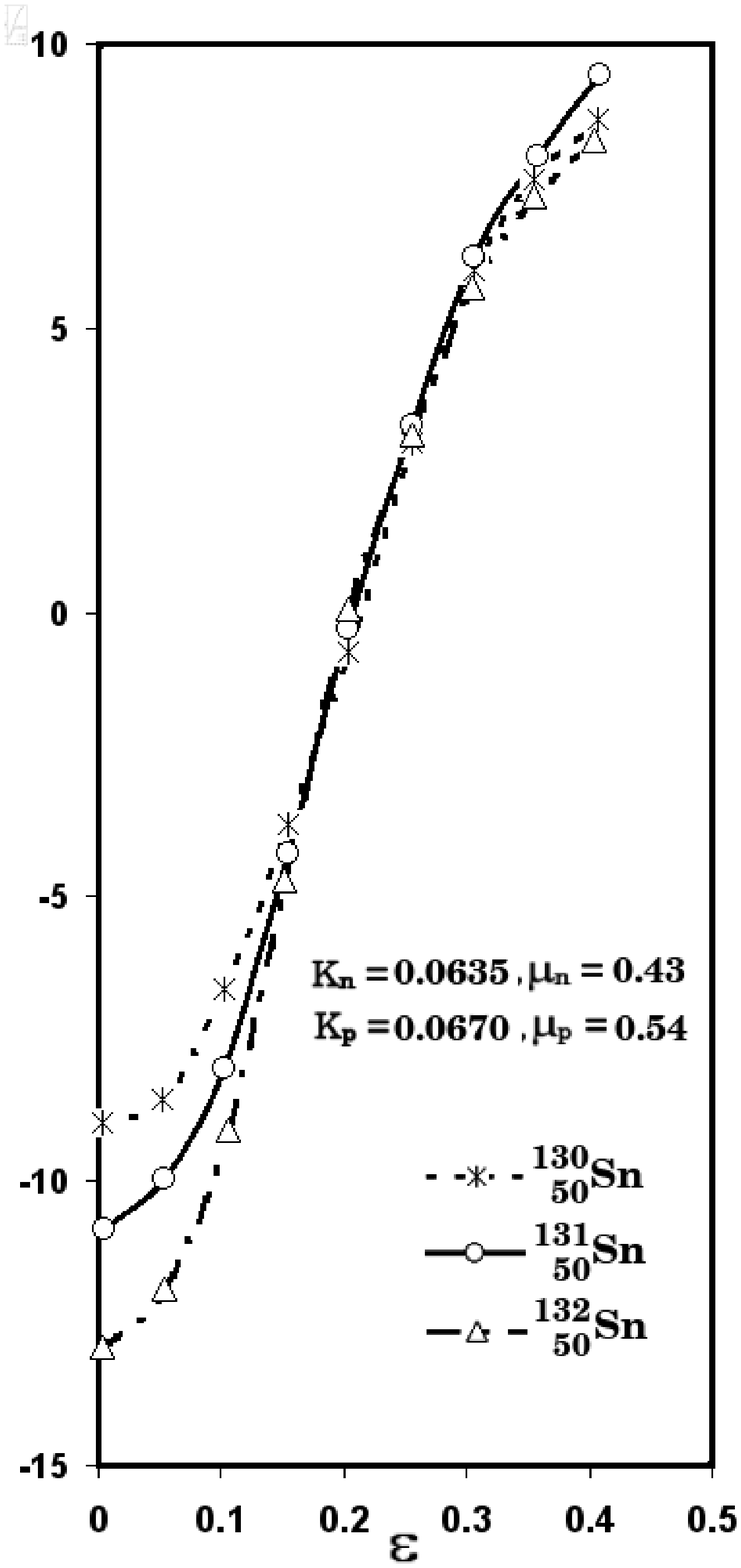}
\caption{Deformation energy for nuclei $^{130-132}$Sn calculated by a drop liquid model with pairing and shell correction~\cite{Myers}. See text.}
\label{fig:130132Sn}
\end{figure}

The total energy of a nucleus is calculated at first approximation by a liquid drop model type ($\widetilde W$), using the mass formula of Myers and Swiatecki~\cite{Myers}. The shell correction ($\delta U$) is calculated by the Strutinsky's method~\cite{Strutinsky}, using Nilsson Hamiltonian~\cite{Quentin}:

\begin{equation}
V_{corr}= -\kappa [\vec{l} \cdot \hat{s} + \mu ({\vec{l}}^2 -<{\vec l}^2>_N)],
\label{eq:NilssonV}
\end{equation}
where $\kappa$ and $\mu$ are the Nilsson's constants.

The pairing correction is calculated using the BCS method~\cite{BCS}. 

Then, the relation for the total energy of the nucleus (Z,N) results:

\begin{equation}
DE(Z,N,\epsilon)= \widetilde W (Z,N, \epsilon) - \widetilde W_S(Z,N) + \delta U_N + \delta U_Z+\delta P_N + \delta P_Z,
\label{eq:DEW}
\end{equation}

where $\widetilde W(Z,N,D)$ is the energy of a nucleus $(Z,N)$ having deformation $D$, and $\widetilde W_S(Z,N)$ the energy in its spherical shape.

The constant of the harmonic oscillator was the suggested by Nilsson~\cite{Nilsson}:

\begin{equation}
\hbar w_0 = 41A^{-1/3}.
\label{eq:Harconst}
\end{equation}

As one said, the total fragments kinetic energy is close to the available energy for light and heavy complementary fragments with masses around $A=104$ and $A=132$, respectively. Let us relate this result to the deformation for nuclei in this mass neighborhood. 

The energies of nuclei $^{106-108}$Mo and $^{106-108}$Tc as a function of their corresponding deformations ($\epsilon$)are presented on Figs. ~\ref{fig:106108Mo} and ~\ref{fig:106108Tc}, respectively. The assumed Nilsson's constants~\etal\cite{Nilsson} for these nuclei are $\kappa_N = 0.678$, $\kappa_P = 0.07$, $\mu_N = 0.33 $ and $\mu_P = 0.35$.

As we can see, those nuclei have a prolate shape with to $\epsilon = 0.3$ in their ground state. If the fragment deformation changes from $\epsilon = 0$ to $\epsilon = 0.3$ the deformation energy will decreases by around 2 MeV, while a change from $\epsilon = 0.3$ to $\epsilon = 0.4$ increases of deformation energy by 4 MeV. This result suggests that these nuclei are prolate and soft between $\epsilon = 0$ to $\epsilon = 0.3$ and became stiff for higher prolate deformations.

The energy as a function of deformation for nuclei $^{130-132}$Sn are presented on Fig. ~\ref{fig:130132Sn}. The assumed Nilsson's constants for these nuclei are $\kappa_N = 0.635$, $\kappa_P = 0.067$, $\mu_N = 0.43$ and 
$\mu_P = 0.54$. One can see that $^{130}$Sn is softer than $^{132}$Sn. For a deformation from  $\epsilon = 0.0$ to $\epsilon = 0.2$ the nucleus $^{130}$Sn spends around 5 MeV while the nucleus $^{132}$Sn, for the same deformation, spends 10 MeV. The neutron number number $N=82$ and proton numbers around $Z=50$ correspond to spherical hard nuclei.

The above characteristics of light fragments,  corresponding to masses from $A = 100$ to $A = 106$, and their complementary fragments, corresponding to mass from $A = 130$ to $A = 132$, makes possible that their maximal values of the total kinetic energy of complementary fragments ($TKE$) are close to the available energy. 

\begin{figure}
\centering
\includegraphics[height=0.5\textwidth]{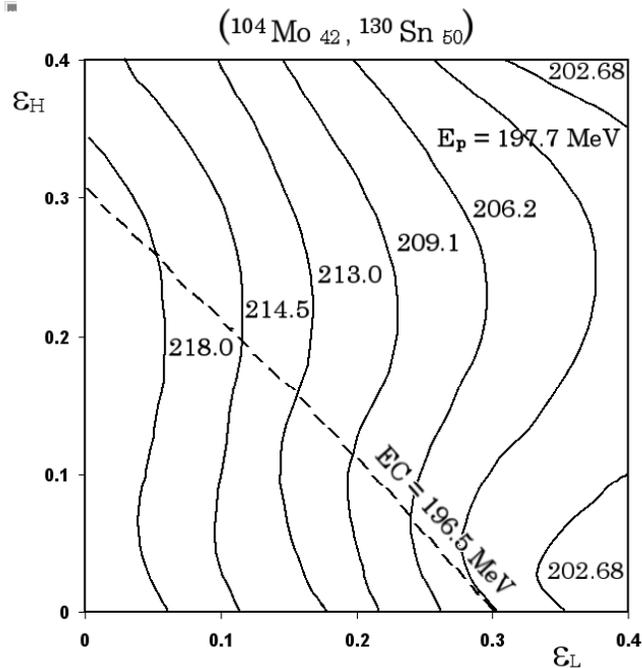}
\caption{Equipotential curves for scission configuration of fragments  $_{42}$Mo$_{62}$, $_{50}$Sn$_{80}$ as a function of their deformation. $\epsilon_L$ and $\epsilon_H$ are the light and heavy fragment deformation. }
\label{fig:EDECScission}
\end{figure}

For the the case of $^{233}$U(n$_{th}$, f), the total kinetic energy of the couple $_{42}$Mo$_{62}$, $_{50}$Sn$_{80}$ is almost equal to the available energy. This results means that the corresponding scission configuration is composed by fragments in their ground state, i.e. $DE = 0.$ for each one.  On the Fig.~\ref{fig:EDECScission} we can see the several equipotential energy of the scission configuration composed by those fragments given by the relation

\begin{equation}
V = CE + DE_{H}+ DE_{L},
\label{eq:VDHE}
\end{equation}

where $\epsilon_H$ and $\epsilon_L$ are the heavy and light fragment deformation energy, respectively, calculated using the Nilsson model\cite{Nilsson} and $CE$ is the interaction Coulomb energy between the two fragments separated by 2 fm. On this curve one obtains that for$\epsilon_H = 0$ and $\epsilon_L=0.3$ the Coulomb energy is equal to the available energy to 204 MeV.

The results are similar to complementary fragments corresponding to the deformed transitional nuclei with $A_L$ between 100 and 106 ($N$ between 60 y 64) and to the spherical nuclei with $A_H$ around 132 ($Z = 50$ y $N = 82$).

For the complementary fragments $_{42}$Mo$_{62}$ and $_{50}$Sn$_{80}$, the maximal value of CE corresponds to ground state nuclei or close to that. This case is unique. Other configurations will need deformation energy, which will be higher for the harder nuclei. On the Fig. ~\ref{fig:130132Sn} is presented the deformation energy for the spherical nuclei $^{130}$Se, $^{131}$Se and $^{132}$Se, respectively. We can see that the double magic nucleus $^{132}$Se need 2 MeV more than $^{130}$Se for going from the spherical state $\epsilon = 0$ to the slightly deformed $\epsilon = 0.05$. The fact that $^{130}$Se is no so hard as $^{132}$Se explain which the highest values of Coulomb interaction energy correspond to values close to the available energy for $^{233}$U(n$_th$,f) as well as for $^{235}$U(n$_th$,f).

\section{Conclusion}

From calculations of scission configurations from thermal neutron induced fission of $^{233}U$ and $^{235}U$,  one can conclude that the highest value of Coulomb interaction energy between complementary fragments correspond to fragmentations ($_{42}$Mo$_{62}$, $_{50}$Sn$_{80}$) and ($_{42}$Mo$_{64}$, $_{50}$Sn$_{80}$, respectively. For both cases the calculated maximal value of Coulomb interaction energy are equal to the available energy of the reaction for spherical ($\epsilon_H=0$) heavy fragments and prolate ($\epsilon_L=0.3$) complementary light fragments, which correspond to their ground states. Moreover the light fragments are soft between $\epsilon_L=0.0$ and $\epsilon_L=0.3$ and hard if they go to more prolate shapes; while the heavy fragment $_{50}$Sn$_{80}$ is no so hard as $_{50}$Sn$_{82}$. The calculated maximal value of Coulomb interaction energy is equal to the measured maximal value of total kinetic energy of fragments. The prescission kinetic energy and intrinsic excitation energy of fragments are assumed to be null. These results suggest that fission process take time to explore all energetically permitted scission configurations.

\end{document}